\documentclass{aa}
\usepackage{mhchem}
\usepackage{natbib}
\usepackage{graphicx}
\usepackage{txfonts}
\usepackage{amsmath}
\usepackage{appendix}
\usepackage{tabularx}

\begin{document}

   \title{Chemical evolution in ices on drifting, planet-forming pebbles}

   \author{Christian Eistrup
          \inst{1}
          \and
          Thomas Henning
          \inst{1}
          }

   \institute{Max Planck Institute for Astronomy (MPIA), K\"onigstuhl 17, Heidelberg, Germany\\
              \email{eistrup@proton.me}}

   \date{\today}

  \abstract
   {Planets and their atmospheres are built from gas and solid material in protoplanetary disks. Recent results suggest that solid material like pebbles may contribute significantly to building up planetary atmospheres. In order to link observed exoplanet atmospheres and their compositions to their formation histories, it is important to understand how icy pebbles may change their composition when they drift radially inwards in disks.}
   {Our goal is to model the compositional evolution of ices on pebbles as they drift in disks, and track how their chemical evolution \emph{en-route} changes the ice composition relative to the ice composition of the pebbles in the region where they grew from micron-sized grains.}
   {A state-of-the-art chemical kinetics code is utilised for modelling chemical evolution. This code accounts for the time-evolving sizes of the solids that drift. Chemical evolution is modelled locally for 0.1 Myr at two starting radii, with the micron-sized solids growing into pebbles simultaneously. The pebbles and local gas, isolated as a parcel, is then exposed to changing physical conditions, intended to mimic the pebbles drifting inwards in the disk midplane, moving to 1 AU on three different timescales. A modelling simplification is that the pebbles are \emph{not} moved through, and exposed to new gas, but stay in the same chemical gas surroundings in all models.}
   {For ice species with initial abundances relative to hydrogen of $>10^{-5}$, such as \ce{H2O}, \ce{CO2}, \ce{CH3OH} and \ce{NH3}, the abundances change by less than 20\% for both radii of origin, and for the two smaller drift timescales (10kyr and 100kyr). For less abundant ice species, and the longest drift timescale (1Myr), the changes are larger. Pebble drift chemistry generally increases the ice abundances of \ce{CO2}, HCN and SO, at the expense of decreasing abundances of other volatile molecules.}
    {}

\maketitle

%

\section{Introduction}

Exoplanets are thought to form in protoplanetary disks around young stars. Exoplanet atmospheres can be observed, chemically characterised, and molecular constituents of the atmospheres can be identified. More generally, carbon-to-oxygen (C/O) ratios of exoplanet atmospheres can be retrieved from directly imaged exoplanets with existing facilities, such as the VLTI \textsc{Gravity}-experiment \citep[e.g., the exoplanets $\beta$ Pic b and HR 8799e, see][]{gravity2020betapic,molliere2020} and the \textsc{IGRINS}-instrument at the Gemini-South Observatory, which was used to retrieve the C/O ratio of the day-side hemisphere of the transiting exoplanet WASP-77Ab \citep[see][]{line2021ctoo}. Such chemical characterisation of exoplanet atmospheres will be accelerated by \emph{JWST}, both in terms of precision (lower uncertainty on C/O ratios) and in terms of the number of exoplanets for which atmospheres will be chemically characterised ($\sim$88)\footnote{Number of unique exoplanets targeted for atmospheric spectroscopic characterisation as described in Public PDFs of ERS, GTO, and Cycle 1 GO programs on Space Telescope Science Institute's \emph{JWST} portal (``Exoplanet''-category).}.

Insights into atmospheric compositions of exoplanets is a window into understanding what the atmospheres, and the exoplanets as a whole, formed from \citep[see e.g.][]{mordasini2016}. The idea is that the elemental chemical compositions of exoplanet atmospheres should reflect the combined elemental compositions of the material(s) that went into forming the exoplanet atmospheres. These are the materials that exist in protoplanetary disks, and which go into forming planets: gas and solids, where solids have both a refractory (rocky, with high desorption temperatures/molecular binding energies) and a volatile (icy, with low desorption temperatures/molecular binding energies) component. The region of protoplanetary disk midplanes where volatile species start to freeze out as ices usually have temperatures below 150K. \ce{H2O}, as the volatile molecule with the highest molecular binding energy \citep[see, e.g.,][from the former of which $E_{\mathrm{bind}}$=5770K was used for this study]{fraser2001,penteado2017}, freezes out below this temperature.

Planetary atmospheres may be formed by run-away accretion of gas surrounding a solid planetary core \citep[as modelled by][with, respectively, without protoplanetary disk chemical evolution]{madhu2014,notsu_eistrup2020}, and accretion of solid bodies moving in disks. There is growing evidence suggesting that micron-sized grains grow during the process of planet formation. When these bodies reach sizes of pebbles, they will drift inwards in the disk \citep[see, e.g.,][]{lambrechts2012,bitsch2019pebblesgiants,trapman2019}. These pebbles are likely to be accreted by forming planets as they drift. Such accretion could lead to the material contained in the pebbles becoming part of the forming planet, and its atmospheres. In other words, the accretion of pebbles by exoplanets could affect the chemical makeup of the planetary atmospheres.

\begin{figure}
    \centering
    \includegraphics[width=0.38\textwidth]{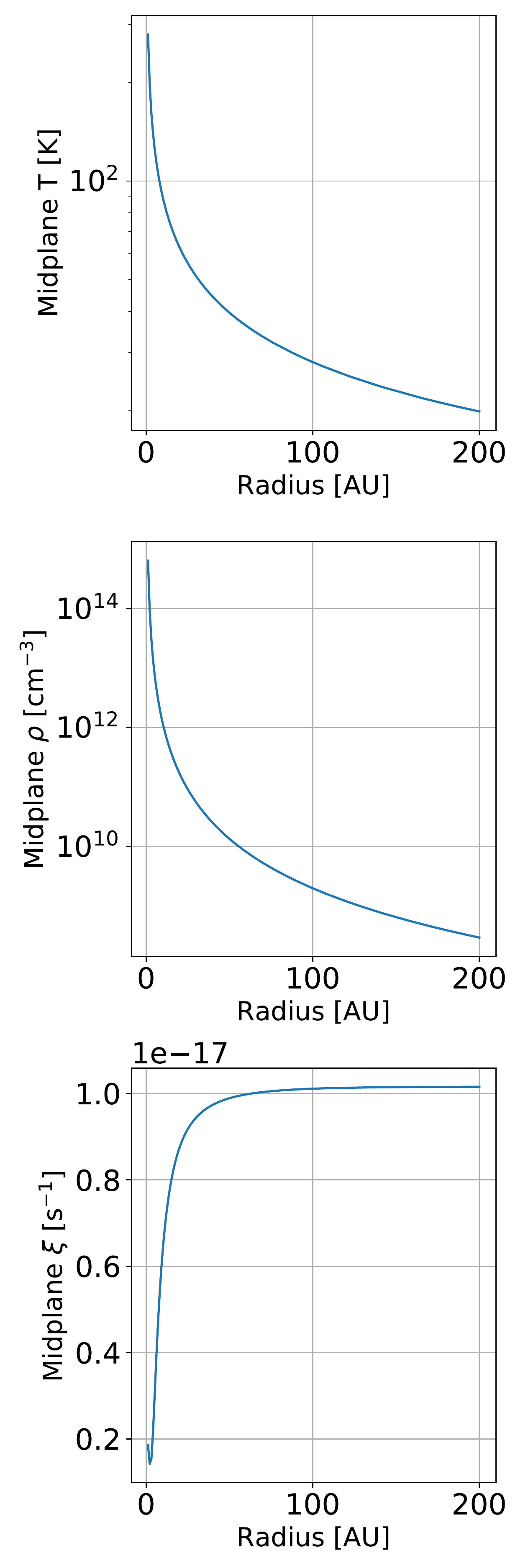}
    \caption{Temperature (top), number density (middle) and ionisation level structures in the disk midplane, as functions of radius. Temperature and density structures are calculated from the prescription for the Minimum Mass Solar Nebula structure in \citet{hayashi1981,aikawa1999}. The ionisation structure is calculated assuming a cosmic ray ionisation rate of $\xi$=$10^{-17}$s$^{-1}$, and accounting for surface density-dependent attenuation of this rate at each radius in the disk midplane, following the prescription from \citet{umebayashi2009}}
    \label{conditions}
\end{figure}
\begin{table*}[]
    \centering
    \begin{tabular}{c|cc|rc}
        Molecule & Initial abundance, wrt H&& $E_{\mathrm{bind}}$ [K]&Radius for final\\
         & 128AU & 200AU & &abundance [AU]\\
         \hline\hline
         &&&\\
        \ce{H2O}& 2.99E-04&3.00E-04 &5770&4.8\\
        \ce{CH3OH}& 4.09E-05&4.37E-05 &4930&4.8\\
        \ce{SO2}&1.23E-09 &4.54E-11 &5330&4.8\\
        \ce{CO2}&6.34E-05 &6.00E-05 &2990&18.2\\
        \ce{NH3}&2.05E-05 &2.11E-05 &3130&18.2\\
        \ce{HCN}&2.70E-07 & 3.48E-07&3610&18.2\\
        \ce{H2S}& 2.68E-06&2.67E-06 &2743&18.9\\
        \ce{SO}&4.64E-08 &1.47E-09 &2600&18.9\\
        \ce{S2}&3.17E-12 &4.89E-11 &2200&31\\
        \ce{CH4}&1.37E-05 &1.83E-05 &1090&-\\
        CO&-&2.21E-05&855&-\\
    \end{tabular}
    \caption{Initial abundances for nine key volatile species, for both radii of drift origin (128AU and 200AU). Fourth column indicates (for reference) the molecular binding energies used for each molecule in the chemical code. Fifth column indicates the radii (outside the iceline radius of a given species) at which the final abundances of these species are evaluated. This set of abundances is not the complete set of species with non-zero initial abundances, but merely the most abundant ice species at the onset of modelling. CO ice is not tracked, but its initial abundance at 200AU (19.5K) is shown, for comparison. The binding energies are adopted from the \textsc{Umist Rate12} \citep{mcelroy13}, as described in \citet{walsh2015}.}
    \label{initabun}
\end{table*}

\begin{table}[]
    \centering
    \begin{tabular}{c|r}
    Element&Relative abundance\\
    \hline\hline
    &\\
         H& 1\\
         He& 9.75$\cdot 10^{-2}$\\
         O& 5.2$\cdot 10^{-4}$\\
         C& 1.8$\cdot 10^{-4}$\\
         N& 6$\cdot 10^{-5}$\\
         S& 6.2$\cdot 10^{-6}$\\
    \end{tabular}
    \caption{Global elemental abundances assumed in this work. All are wrt H$_{\mathrm{nuc}}$.}
    \label{elementals}
\end{table}
In order to establish a link between the chemical composition of a planetary atmosphere, and the formation history of the planet and its atmosphere, it is important to constrain the chemical composition of the volatile ices carried on these pebbles, as they drift from the outer, colder regions of the disk, inwards. The volatile ices are of special importance, since these ice components are the main carriers of carbon, oxygen, nitrogen and sulphur, which go into forming the gas-phase phase species constituting exoplanet atmospheres. Pebble drift in protoplanetary disks is an active area of research \citep[e.g.,][]{brauer2008,birnstiel2012,birnstiel2015}, including the accretion of pebbles onto planets \citep{mordasini2016,bitsch2015,bitsch2016,bitschtrifonov2020,bitschbattistini2020,cridland2016,cridland2017}, and how this accretion affects the resulting compositions of the exoplanet atmospheres \citep[see, e.g.][]{madhu2014,drummond2019}.

However, most of these efforts have, thus far, assumed a pre-defined ice abundance for the pebbles, which remains unchanged from the beginning of simulations until the pebbles are accreted onto the forming planet. An exception is the desorption of volatile ices, which is assumed to happen when the disk temperature around the pebbles reaches a certain level, corresponding to an icy species desorbing into the gas-phase. The work by \citet{boothilee2019}, did utilise a chemical kinetics network through the \textsc{Krome}-package \citep{grassi2014} to model chemical evolution on drifting solids, although they only considered hydrogenation reactions for ices, and not two-body ice reactions via the Langmuir-Hinchelwood-mechanism \citep[see, e.g.][]{hasegawa1993,herbst2021review}. Lastly, the work by \citet{cevallossoto2022} coupled disk gas and dust evolution with chemical evolution, including pebble drift. Their results will be discussed in Section \ref{disccomp}.

This paper investigates how the chemical evolution in the volatile ices on growing grains, and the subsequent chemical evolution of pebbles moving through a disk midplane changes the chemical ice composition of the pebbles. The volatile ices will be the focus, as these are the dominant carriers of elemental carbon and oxygen, apart from carbonaceous grains and silicates. The investigation will aid a better understanding of the actual icy compositions of these pebbles as they get accreted onto a forming planet, and will, in turn, provide a better basis of making the connection between characterised exoplanet atmosphere and protoplanetary disk midplane evolution. This will improve the framework that is needed when \emph{JWST} results of characterised exoplanet atmospheres will be put into the context of how these exoplanets originally formed in their natal protoplanetary disks.

\section{Methods}
\label{methods}

This paper utilises the \textsc{Walsh} chemical kinetics-code \citep[used in][]{walsh2015,eistrup2016,eistrup2018}, which is based on the \textsc{UMIST Rate12} network \citep{mcelroy13}, and includes gas-phase reactions, gas-grain interactions, grain-surface reactions (including the Langmuir-Hinchelwood mechanism), and accounts for grain growth \citep[following][]{eistrup2022graingrowth}. The focus here is the midplane of a protoplanetary disk. This region is considered well-shielded from radiation from the central star. However, cosmic rays (CRs) are assumed to be able to penetrate through the upper layers of the disk, and, with surface density-dependent attenuation length of $\lambda_{\mathrm{CR}}$=$ 96\frac{\mathrm{g}}{\mathrm{cm}^2}$ \citep[see][]{umebayashi1981,umebayashi2009}, these CRs are able to ionise \ce{H2} and He, which subsequently produces a secondary UV-field.

These UV photons can then photo-dissociate species in the gas phase and in the ice, with reaction rate coefficients $k_{\mathrm{pd}}$ in both phases calculated using:

\begin{equation}
    k_{\mathrm{pd}}=\alpha \left(\frac{T}{300\mathrm{K}}\right)^{\beta}\frac{\gamma}{1-\omega} \mathrm{s}^{-1},
\end{equation}
where $\alpha$ is the CR ionisation rate (surface-density-attenuated) seen in the bottom plot of Fig. \ref{conditions}, $\beta$ accounts for the fact that some species are photo-dissociated by discrete line absorption \citep[see][]{gredel1987,woodall2007,chapparomolano2012}, $\gamma$ is the efficiency of the cosmic-ray ionisation events as defined in Eq. (8) of \citet{gredel1989}, and $\omega$ is the albedo of dust grains in the far-UV.
For two-body, grain-surface reactions, a ratio of diffusion-to-binding-energies $\frac{E_\mathrm{{diff}}}{E_\mathrm{{bind}}}$=0.5 is assumed. Two monolayers of molecules are assumed to be chemically active in the grain surface ices. Furthermore, for grain-surface hydrogenation reaction, the barrier width for quantum tunneling between surface sites is set to $q_\mathrm{{qt}}$=1$\AA$. Both these assumption are motivated by \citet{cuppen2017review}.

The \textsc{Walsh} code is used to track the chemical evolution of pebbles in a protoplanetary disk midplane. The chemical code considers two different physical effects, as described below:

\begin{enumerate}
    \item Grain growth is assumed to take place locally in the disk midplane, at two specific radii (128 and 200 AU), with constant local temperatures of 25 and 19.5K. The two radii are selected to represent different starting locations of pebbles migration, with associated differences in starting temperatures (at 25K CO is in the gas phase, whereas at 19.5K, it is frozen out as ice). The grain growth follows the models by \citet{krijtbosman2020}, which tracks the total available grain surface area of the grain population, along with the number of grains of different sizes. For each time step in the evolution of this grain population, the chemical code is run to track the chemical evolution, as it depends on temperature, density, ionisation level and available grain surface area. 

    The chemical evolution is run alongside grain growth for 0.1Myr, locally at both radii. This represents the time it takes to grow grains from micron-sizes, where the grains are well-coupled with the local gas motion, to mm-sizes, at which point the grains are assumed to dynamically decouple from the gas, lose orbital angular momentum to the gas, and hence start drifting inwards in the disk \citep[see, e.g.,][]{krijtbosman2020}.

    \item After 0.1 Myr of grain growth and chemical evolution at each radius, all solid material is assigned the ice composition resulting from the local chemical evolution (see Table \ref{initabun}), and all solids are assumed to be compact, spherical pebbles of radius $R$=0.6mm, with uniform chemical compositions. These pebbles are subsequently assumed to drift inwards through the midplane at constant speed. In reality, it is predominantly pebbles that drift (solid bodies with Stokes numbers $\sim$1), and neither smaller grains (that are dynamically coupled to the gas), nor larger solids (which are dynamically independent of the gas) are expected to drift as efficiently as pebbles. However, in this work, it is assumed that upon reaching 0.1 Myr of local chemical evolution and grain growth at a given radius, all grain mass in the grain population is carried in 0.6mm sized spherical pebbles.

    Following Eq. 2 from \citet{birnstiel2012}, a silicate volume density for the pebbles (3g/cm$^{3}$), and a MMSN surface density structure\footnote{$\Sigma_{\mathrm{gas}}=1700\left(\frac{r}{1AU}\right)^{-1.5}$g/cm$^{2}$} from Eq. 2.6 in \citet{hayashi1981}, a pebble size of 0.6mm corresponds to Stokes numbers St=0.24 (at disk midplane radius $R$=128AU) and St=0.47 (at $R$=200AU). Using these Stokes numbers, the radial drift speed $v_{r}$ of the pebble drifting from 128AU or 200AU can be estimated using Eqs. 17 and 18 from \citet{brauer2008}:

    \begin{equation}
        v_(r)=-\frac{2v_{n}}{\mathrm{St}+\frac{1}{\mathrm{St}}}= -\frac{2 \frac{c_{s}^2}{2V_{K}}(\delta+7/4)}{\mathrm{St}+\frac{1}{\mathrm{St}}},
    \label{drift_speed}
    \end{equation}

    where $c_{s}=\frac{kT_\mathrm{{gas}}}{\mu}$ is the isothermal sound speed for a given midplane gas temperature, $T_\mathrm{{gas}}$, and mean molecular weight, $\mu$ (taken here to be $\mu=2.3m_\mathrm{{proton}}$ for a gas mainly composed of H$_2$ and He), and $V_{K}=\Omega_{K}r=\sqrt{\frac{GM_{\star}}{r^{3}}}$ is the Keplarian velocity, with the gravitational constant $G$, the central stellar mass $M_{\star}=M_{\odot}$, and $r$ being the disk midplane radius. $\delta$ in Eq. \ref{drift_speed} is the power-law index in the assumed disk surface density profile \citep[from Eq. 2.6]{hayashi1981}, taken to be 1.5 in this work.

    Based on these assumption, $T_\mathrm{{gas}}$=25K at $r$=128AU and $T_\mathrm{{gas}}$=20K at $r$=200AU, the drift speeds for the 0.6mm-sized pebble is 25m/s at 128AU and 43m/s at 200AU. Taking these drift speeds to be constant throughout drift, the drift time scales $\tau_{d}$ to get from 128AU to 1AU is $\tau_{d}$(128AU)=2.4$\cdot 10^{4}$ years, and $\tau_{d}$(200AU)=2.2$\cdot 10^{4}$ years. If the size of the pebble corresponded to at Stokes number St=1 at $r$=200AU, then the drift speed would be $v_{r}$=55m/s and the drift timescale would be $\tau_{d}$=1.7$\cdot 10^{4}$ years. This is therefore taken to be the minimum drift timescale. Several factors, including pressure bumps, may cause the actual timescale to be longer than this minimum. As such, the short drift timescale calculated above is derived assuming a smooth disk. This is further discussed in Section \ref{model_caveats}.

    From these drift speeds and drift timescales, three different bracket choices are made for the drift speeds of the pebbles. These choices capture the minimum drift timescale, as well as two longer drift timescales, and are motivated by \citet{brauer2008,birnstiel2012,birnstiel2014, birnstiel2015}:
    \begin{enumerate}
        \item Slow, where the drift time from origin (128AU or 200U) to 1AU is 1Myr.
        \item Medium, where the drift time from origin to 1AU is 100kyr.
        \item Fast, where the drift time from origin to 1AU is 10kyr.

    \end{enumerate}
    \end{enumerate}
    Pebbles drifting from both 128AU and 200AU were assumed to be of size $R$=0.6mm (and each starting location features a unique ice composition for the pebbles, see Table \ref{initabun}). This was the grain size assigned to all grains when running the chemical code during radial drift. In order to conserve the total grain mass (which was assumed constant), the grain number density with respect to H nuclei, $N_{\mathrm{grain}}$, is scaled down in accordance with how much mass is carried in a 0.1$\mu$m (spherical) grain versus how much is carried in a $R_{\mathrm{adopt}}$=0.6mm pebbles, giving us an adopted grain number density per H nuclei:

    \begin{equation}
        N_{\mathrm{grain, adopt}}=\left(\frac{0.1\mu\mathrm{m}}{0.6{\mathrm{mm}}}\right)^{3}\times 1.3\times 10^{-12}\mathrm{H}^{-1}=6\times10^{-24} \mathrm{H}^{-1},
    \end{equation}
    Where $1.3\times 10^{-12}$ is the grain number density with respect to H nuclei, assuming that all grains are made of silicates, are 0.1$\mu$m in radius, and a gas-to-dust ratio of 100.

    To ease the implementation of the grain drift into the chemical code, the treatment of the chemical evolution of the drifting solids is simplified. While in reality drifting solid bodies move through different physical environments (with increasing temperature and density, which is included here), the drifting solids likely also move through gas and icy solids in the disk midplane that have experienced different types or degrees of chemical evolution than have the solids drifting from further out in the disk. For this reason, it is reasonable to assume that an ice-covered drifting solid body will experience, not only increasing temperatures and densities, but also different chemical compositions of the surrounding gas than the drifting body experienced further out. Furthermore, when crossing icelines of different molecules, it is likely that the gas density inside the iceline will be higher and have a higher abundance of the molecule associated with the iceline, stemming from the desorption of that given molecule off the grain surfaces of solids, which have drifted across the iceline and deposited this molecule as gas, prior to the drifting of the solid bodies that are tracked here.

    Here, this situation is simplified by assuming the following: the material that the models start out with stays the same throughout the chemical evolution. That is to say, the material is considered a closed parcel, where no material is added or lost. Within this parcel, solids and gas coexist, chemical reactions can take place between volatile chemical species, and molecules can freeze out onto solid surfaces, or desorb off the surfaces. However, the sum total of elemental C, O, N and S, and the total mass of solids stay constant (see Table \ref{elementals}). In order to mimic the effects of drift on the chemical composition of the solids in the parcel, the parcel of gas and solids is exposed to changes in physical environment (by changing temperature, particle density and ionisation level) according to the different physical conditions found further in in the disk midplane, see Fig. \ref{conditions}. The temperature and density structures in the disk midplane are calculated following the prescriptions by \citet{umebayashi2009}. The ionisation structure is calculated assuming a cosmic ray ionisation rate of $\xi$=$10^{-17}$s$^{-1}$, and then accounting for surface density-dependent attenuation of this ionisation rate at each radius in the disk midplane, following the prescription from \citet{umebayashi2009}.

    This simplified treatment of pebbles drifting in a disk midplane is justified by the fact that increasing temperatures likely lead to less ice chemistry, and accretion of gas molecules on the solid body surfaces, than would be the case assuming constant or decreasing temperatures. Furthermore, some of the timescales for drift considered here are short (10kyr and 100kyr) in comparison with timescales for ice chemistry on grains \citep[see, e.g.,][]{eistrup2018,eistrup2019_comet_moxy}, so the effect of the solid bodies drifting through gas of varying compositions in reality may be negligible for the composition of the ices of these solids. However, modelling assuming ice-covered pebbles to be moving through varying compositions of gas corresponding to different regions of the disk midplane, should be pursued in future work. This may be especially interesting if assuming that ice species on pebbles that drift early desorb into the gas phase in the inner disk, thereby moving material from ice in the outer disk to the gas in the inner disk \citep[see,][and discussion hereof in Sec. \ref{disccomp}]{cevallossoto2022}. Pebbles that drift later may, therefore, in reality, drift through regions with higher gas densities than early-drifting pebbles, and this may increase the efficiency of gas-grain interactions for gas-phase molecules that collide with, and stick to, the surface of a grain.

\section{Results}
Six models were computed, for each of the six permutations of the model setup:
\begin{itemize}
    \item Drifting starting at 128AU or 200AU.
    \item Drifting to 1AU taking 10kyr, 100kyr or 1Myr.
\end{itemize}

\begin{figure*}
    \centering
    \includegraphics[width=0.98\textwidth]{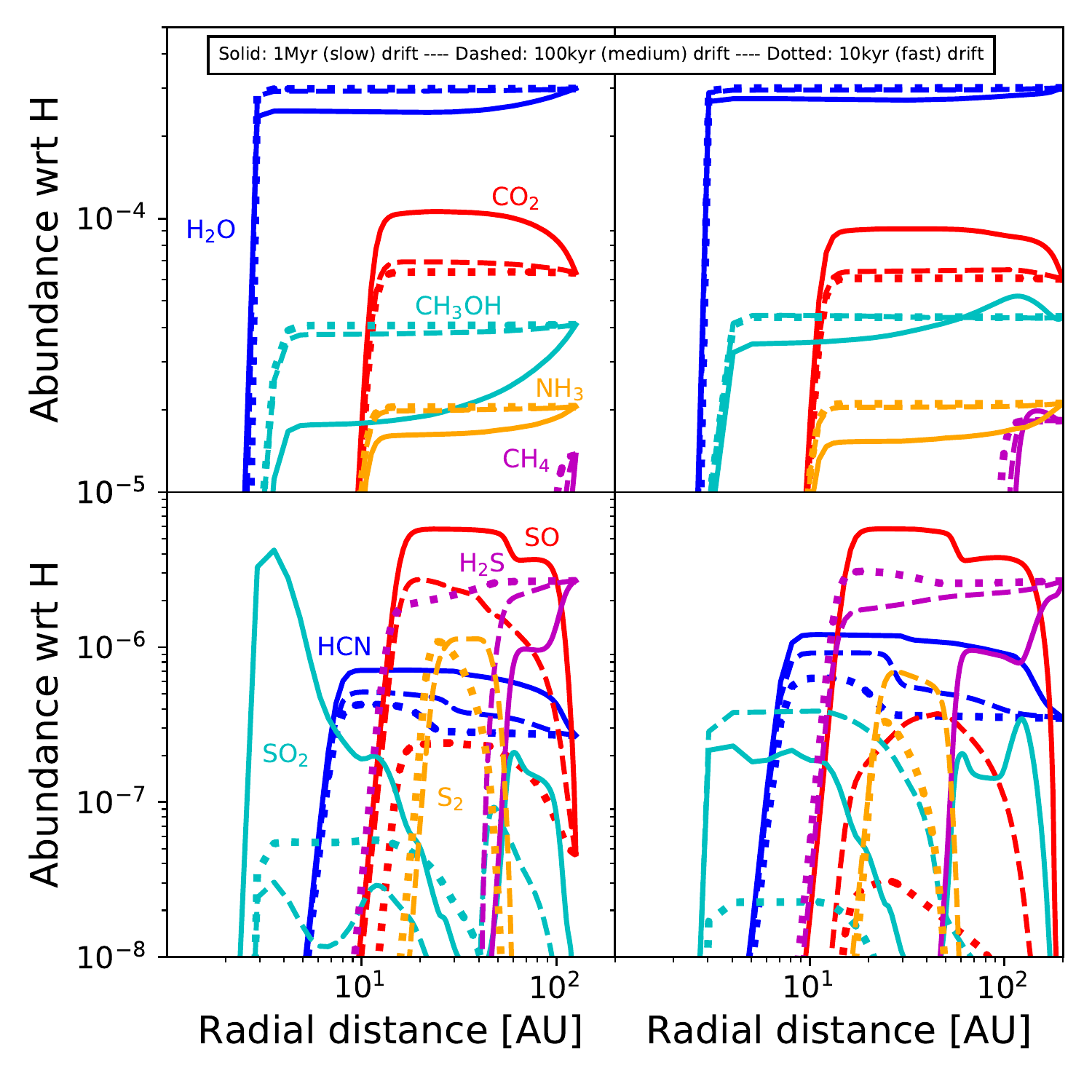}
    \caption{Abundances of volatile ices on a drifting pebble as a function of radius (time), where the pebbles moves inwards with time. Left is drift from 128AU, right is drift from 200AU. Top panels are for species with abundances larger than $10^{-5}$ wrt H (\ce{H2O}, \ce{CO2} \ce{CH3OH}, \ce{NH3} and \ce{CH4}). Bottom panels are for species with abundances ranging from $10^{-8}-10^{-5}$ (\ce{HCN}, \ce{SO}, \ce{SO2}, \ce{H2S} and \ce{S2}). Solid profiles are for slow drift (1Myr), dashed profiles are for medium drift (100kyr) and dotted profiles are for fast drift (10kyr). Colors correspond to the same species in both panels. Note that top and bottom panels have different dynamical $y$-axis ranges.}
    \label{drifting}
\end{figure*}

The model results are show in Figure \ref{drifting}. The left-hand panel of the figure shows drifting from 128AU, whereas the right-hand panel is drifting from 200AU. 10 chemical species are tracked from 128AU (\ce{H2O}, \ce{CO2} \ce{CH3OH}, \ce{NH3}, \ce{HCN}, \ce{CH4}, \ce{SO}, \ce{SO2}, \ce{H2S} and \ce{S2}), whereas 11 are tracked from 200AU (CO additionally). The reason for excluding CO ice from the tracking starting at 128AU is that the temperature there increases to 25K, which is above the iceline temperature of CO, so throughout the tracking of solids drifting from 128AU, CO is mostly in the gas phase, and not ice on the drifting solids.

The solid profiles in both panels are for slow drift (1Myr), the dashed profiles are for medium drift speed (100kyr), and the dotted profiles are for fast drift(10kyr). Following the profile for an ice species from right-to-left in each panel corresponds to following the drifting pebbles as they move inwards at constant radial speed through the disk midplane. It is seen in both top panels that the 4 most abundant ice species (\ce{H2O}, \ce{CO2} \ce{CH3OH}, \ce{NH3}, and CO in the right-hand panel) do not change abundances much when comparing the slow drift to the fast drift. For ice species less abundant than $10^{-5}$ (\ce{SO}, \ce{SO2}, \ce{H2S}, \ce{S2} and HCN), the changes in abundances across drifting speeds are larger, which reflects the fact that these species are inherently less abundant than the other 4 species, and are therefore more affected by small changes to the abundances of the other 4 species.

For all species, it is seen that encountering their icelines, they do not transfer from gas to ice instantly, but instead a gradual decrease in ice abundance is seen going inwards. This is a consequence of icelines being treated here, not as a fixed transition radius (with temperature), but as a balance between the effects of freeze-out (for gas molecules sticking onto grain surfaces as ice) and desorption (for ice molecule that desorb into the gas phase). These two effects are at play for all physical conditions, but under colder conditions the freeze-out term in the chemical kinetics code will dominate and most molecules of a given species will exist as ice on grain surfaces, and under warmer condition the desorption term will dominate, and most of the molecules will exist in the gas-phase. Around the iceline of a species, the terms will be equal, and the molecules of a given species will co-exist, half in the gas, half in the ice. See, e.g., \citet{eistrup2018} for the behavior of the gas abundances in relation to the gradient of the ice abundances as shown here.

\begin{figure*}
    \centering
    \includegraphics[width=\textwidth]{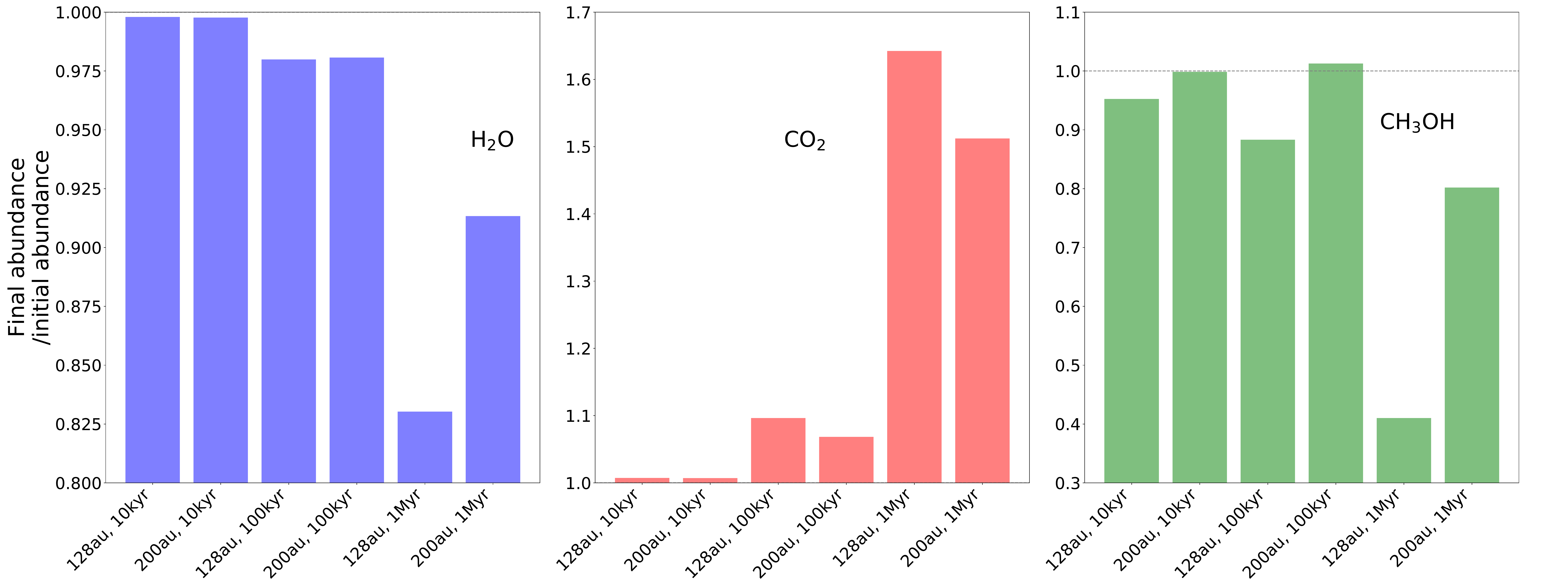}\\
    \includegraphics[width=\textwidth]{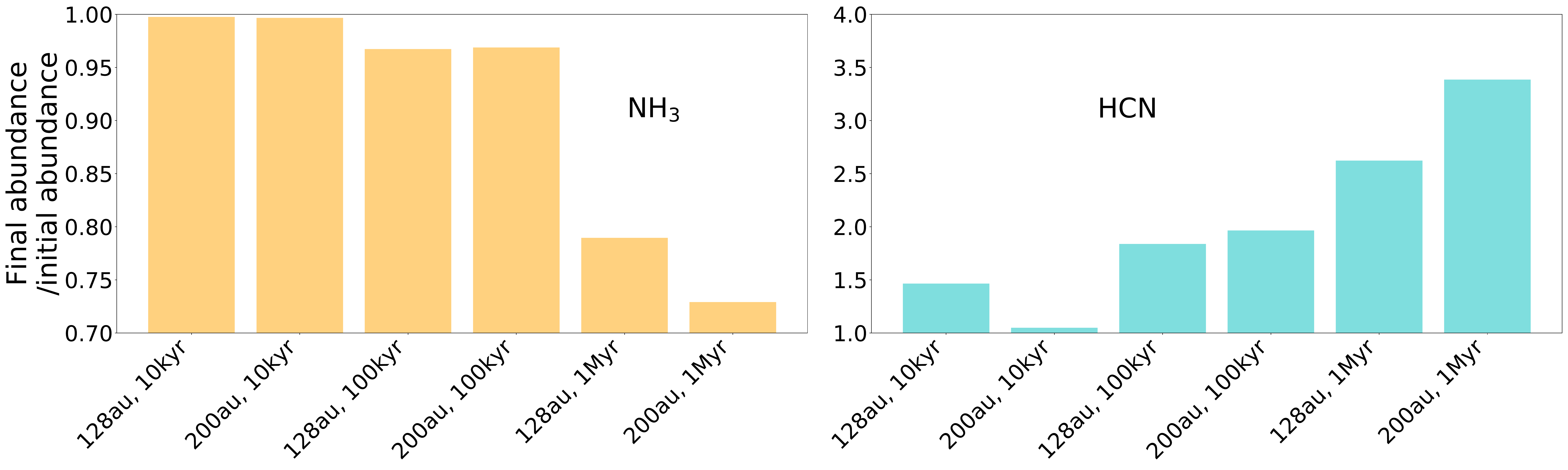}\\
    \includegraphics[width=\textwidth]{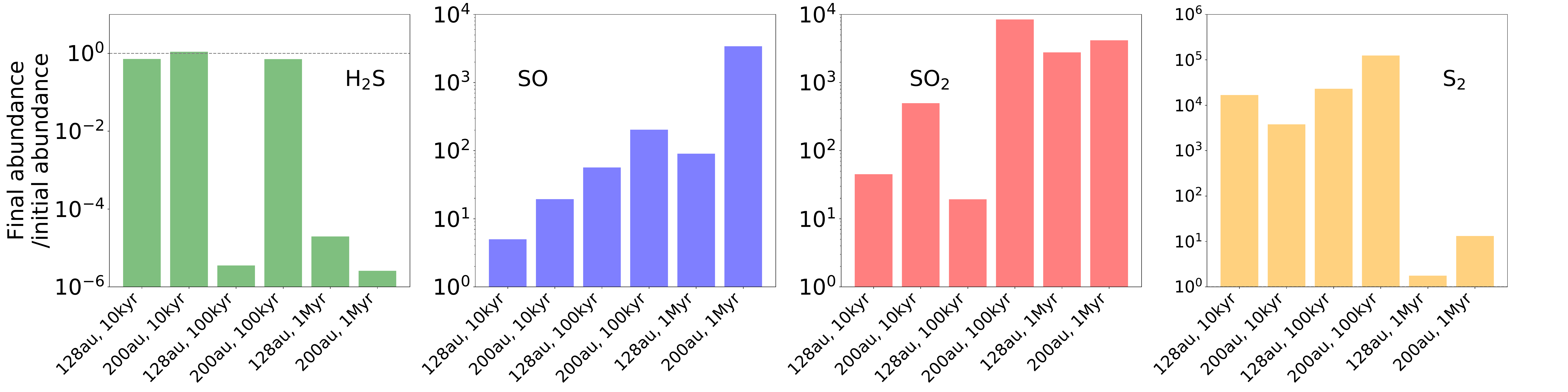}\\
    \caption{Bar plots corresponding to ratios of final abundances of different species relative to initial abundance at the onset of drifting. Initial abundances are seen in columns two and three in Table \ref{initabun}, and the radii at which final abundances are read out of the simulation are seen in column four in Table \ref{initabun}. The $y$axis indicates the ratios of these values, for the six different permutations of modelling setup (on the $x$-axis). Drift time increases (drift becomes slower) going from left to right in each panel. Note that species encountering their icelines further out (see Fig. \ref{drifting}), will experience less drifting time to change their initial abundances than will species with icelines further in. Top and middle rows have linear $y$-axes, bottom row has logarithmic $y$-axis. Note that all $y$-axes feature different dynamical scales.}
    \label{relvals}
\end{figure*}

In order to gain a better impression of the difference that it makes to drift from 128AU versus 200AU, and the difference between slow and fast drift, comparisons between the initial abundances of the solids, and the abundances of the solids just before the drift across the iceline of different ice species are shown in Figure \ref{relvals}. The $y$-axes for the top panels in this figure is linear, whereas it is logarithmic for the bottom panels. The values of the $y$-axes are the ratios of final abundances of each molecule (taken before the solid has crossed the iceline of given ice species, see fourth column in Table \ref{initabun}) to the initial abundances of each molecule (which are the initial abundance values at the beginning of the drifting). The initial abundances of the ice species are different, depending on the initial drift location, see Table \ref{initabun}. For each of the eight ice species featured in Figure \ref{relvals}, the fourth column in Table \ref{initabun} indicates which radii (same for both drifting speeds) the final abundances of these species are being adopted from.

The $x$-axis categories are the same for all 9 species. The first three bars for each species are for fast drift (10kyr), medium drift (100kyr) and slow drift (1Myr). The first and third, and fifth bars are for drifting from 128AU, and the second, fourth and sixth bars are for drifting from 200AU. For \ce{CH3OH} ice in the top row, and \ce{H2S} in the bottom row, a horizontal line indicates a $y$-axis-value of 1, corresponding to a net zero change in abundance of the species during the drifting.

For \ce{H2O} ice it is seen that drifting, no matter the speed or origin, causes a decrease in the abundance. The decrease is smaller (<0.4\%) for faster drift than for medium drift (<2\%) and slow drift ($\sim$10-20\%). For \ce{CO2}, the opposite is seen: a small increase (<1\%) for fast drift, a larger increase (<10\%) for medium drift and a more significant increase for low drift of 50\% when drifting from 200AU, and 65\% when drifting from 128AU. This trend, either larger increase/decrease in abundances with \emph{slowing} drift speed is visible for all species except \ce{CH3OH} and \ce{SO2} ice, suggesting that the faster the drift, the less changes to the initial abundances. This is logical, since faster drift allows less time for chemical processing in the ices, than does slower drift. For the species in the top row, the changes in ice abundances for fast drift is <5\%. For medium drift (100kyr), the changes in abundances range from <2\% for \ce{H2O} ice and <5\% for \ce{NH3} ice, to <8\% for \ce{CO2} ice and less than 15\% for HCN ice and \ce{CH3OH} ice. \ce{H2O} ice and \ce{NH3} ice both decrease in abundance for all drifting scenarios, as compared with the initial abundances, whereas HCN ice and \ce{CO2} ice increase in abundance for all scenarios. \ce{CH3OH} ice decreases in abundance for all scenarios, except for slow drift originating at 200AU, in which case there is a $\sim5\%$ increase in abundance.

For sulphur-bearing species in the bottom row, it is generally seen that SO ice and \ce{SO2} ice increase in abundances for all drift speeds (and \ce{S2} ice for fast and medium drifts), but most for slow drift (albeit from low initial abundances). The elemental oxygen for this increase is sourced from \ce{H2O}, which decreases in abundance for all drifts, except fast drift from 200AU (in which case it increases in abundance by $\sim10\%$).

\section{Discussion}\label{disc}

This section puts the results from the previous section into the context of the astrophysical process of planet formation. This section will also discuss the shortcomings of the method used, and propose an improved framework to be developed in the future.

\subsection{The matter of drift speed}

This paper has utilised assumptions for drift speeds that vary by factors of up to 100. While realistic drift speeds have not been derived analytically or numerically from first principles, the speeds used are motivated by a range of drift speeds found in the literature \citep{birnstiel2012,birnstiel2015}. The aim was thus not to provide a precise prediction to a specific pebble drift scenario, but rather to explore and compare the chemical changes that happens when assuming three different representative drift speeds.

It is clear from Fig. \ref{drifting} that the changes to all abundances as a function of radius (time) is monotonous with changing drift speeds. In addition to this, it is also clear from the panels in Fig. \ref{relvals} that the longer (drift) time is assumed, the larger changes to volatile abundances can be caused by chemical evolution. That means that assumption of either slower or faster drifts than assumed here can be, at least qualitatively extrapolated from the results presented here. However, it is noted that for the four most abundant species (\ce{H2O}, \ce{CO2}, \ce{CH3OH} and \ce{NH3}), the abundance changes amount to $<70\%$ of their initial abundances overall, and for fast and medium drift speeds, these changes are all $<20\%$. These changes are thus smaller than the relative ratios of the different species individually. In other words, and as is also seen in Fig. \ref{drifting}, the above mentioned four species which are top four in the initial abundance hierarchy, do not switch places individually in this hierarchy.

Lastly, it is noted that the difference in abundances across drift scenarios for the four most abundant volatile ices is significantly larger for the slow drift scenarios, relative to the medium and fast drift scenarios. This is explained by the timescale of chemical evolution in ice in disk midplanes, which is usually $>100$kyr. Thereby, it is only in the case of slow drift that significant abundance changes are seen, whereas the differences between fast and medium drifts are comparatively smaller.

\begin{figure*}
    \centering
    \includegraphics[width=1\textwidth]{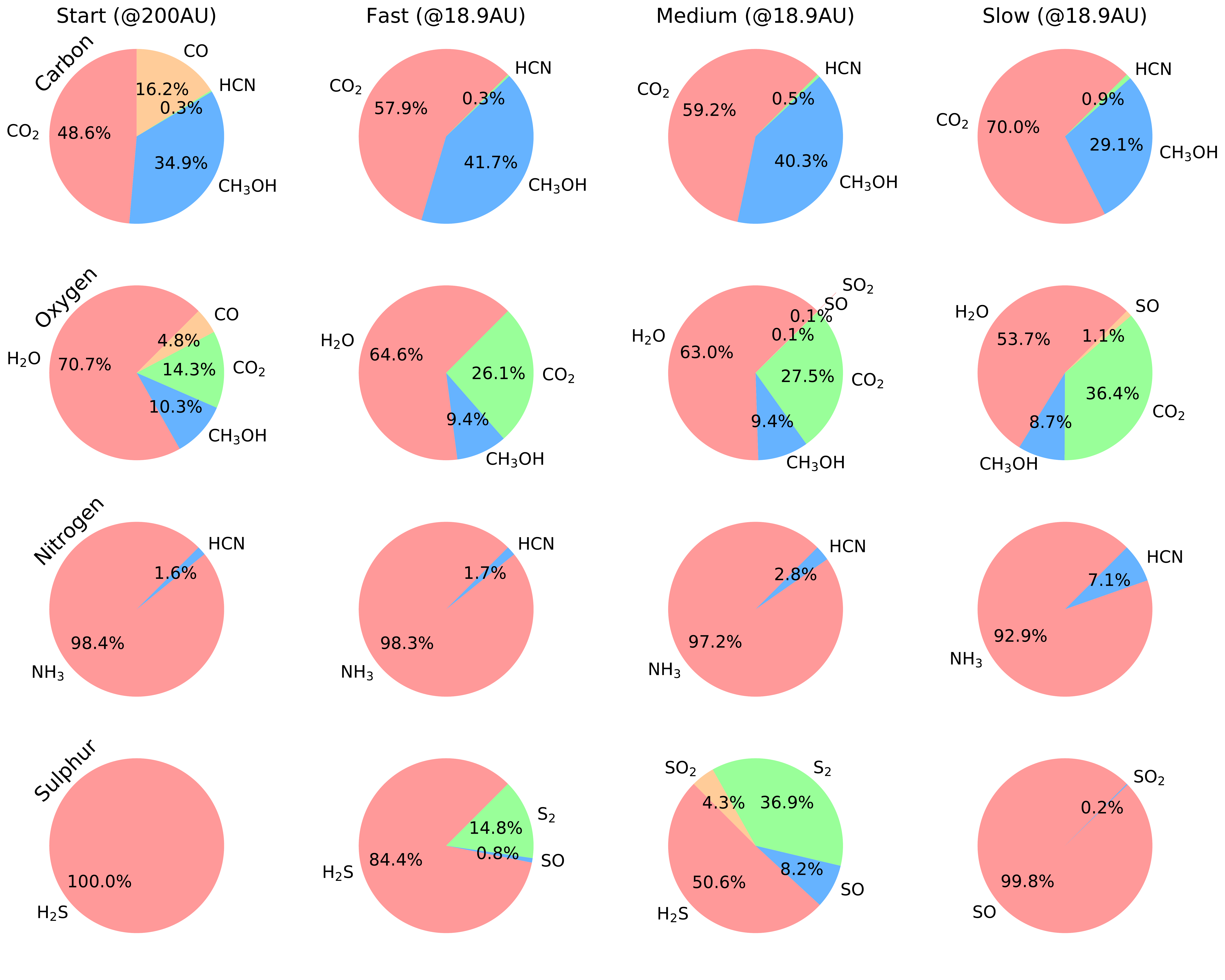}
    \caption{Pie charts showing the percentage of each chemical element (from top to bottom: carbon, oxygen, nitrogen and sulphur) that is carried in different volatile ices (the abundances of \ce{CO2} and \ce{SO2} were doubled when accounting for the oxygen budget in the second row, and the abundance of \ce{S2} was likewise doubled when accounting for the sulphur budget in the fourth row). Left column features the starting abundances for pebble drift starting at 200AU. The second, third and fourth columns represent fast (10kyr), medium (100kyr) and slow drift (1Myr), respectively. Going from left to right therefore corresponds to increasing the chemical timescales. The pie charts in these columns represent pebble ice compositions at 18.9AU, which is the innermost midplane radius at which none of the shown volatile molecules have desorbed yet (see Table \ref{initabun}), except for CO ice, which is only present in the ices at the beginning of drifting (see ``Carbon'' and ``Oxygen'' pie charts in the first column).}
    \label{pies}
\end{figure*}

\subsection{Changes for main volatiles}
\label{}

As seen in the top left panel in Fig. \ref{relvals}, \ce{H2O} changes only by less than 20\% for all drifting scenarios. It is, however, seen that all drifting scenarios cause a decrease in the abundance of \ce{H2O} ice, which is caused by the photodissociation of \ce{H2O} ice by CR-induced photons via the reaction
\begin{equation}
    \ce{iH2O ->[\gamma_{\mathrm{CR}}] iOH + iH}
\end{equation}
Here, ``i'' denotes an ice species. A generally decreasing trend (except for medium drift speeds originating at 200AU) is seen for \ce{CH3OH} ice, which is also caused by CR-induced photodissociation:
\begin{equation}
    \ce{iCH3OH ->[\gamma_{\mathrm{CR}}] iOH + iCH3},
\label{methanol}
\end{equation}
and the elemental oxygen and carbon lost from these two species are mainly processed into \ce{CO2} ice, which sees an increase in abundance for all drifting scenarios, via the two-body grain-surface reaction:
\begin{equation}
    \ce{iOH + iCO -> iCO2 + iH}
\end{equation}
Ice chemistry during drift therefore processes elemental oxygen carried in OH from \ce{H2O} ice and \ce{CH3OH} ice into \ce{CO2} ice, with interaction with the sticking CO from the surrounding gas being the source of elemental carbon in the form of CO that goes into \ce{CO2} ice. The \ce{iCH4} produced via reaction \ref{methanol} further undergoes hydrogenation to form \ce{CH4} on the grain surface, which, due to its low binding energy of $E_{\mathrm{bind}}$=1090K (see Table \ref{initabun}), desorbs into the gas, when the midplane gas temperature is $T_{\mathrm{gas}}$>25K. These effects have been previously modelled and described in, e.g., \citet{eistrup2016,eistrup2018,bosman2018}. An implication of this is that the drifting of any one pebble may lead to decreasing CO gas abundances in the midplane, not just at one radius, but along the entire radial drift trajectory that the pebble follows, as long as it remains outside the icelines of \ce{H2O}, \ce{CH3OH} and \ce{CO2}. Furthermore, the dissociation of \ce{CH3OH} ice leading to the production of \ce{CH4} gas may enhance the \ce{CH4} gas abundance along the pebble's drift trajectory, although the amount of \ce{CH4} gas produced depends of the initial assumption for the abundance of \ce{CH3OH} ice.

For nitrogen, the abundance of \ce{NH3} ice decreases for all drifting scenarios, due to CR-induced photodissociation producing \ce{NH2} and NH. The elemental nitrogen in these species is then processed into \ce{N2} which desorbs into the gas-phase for $T_{\mathrm{gas}}$>19K, and HCN ice, which sees an increase in abundance for all drift scenarios. For sulphur, \ce{H2S} decreases for all drift scenarios, and SO, \ce{SO2} and \ce{S2} all increase. Interestingly, SO ice is the dominant carrier of sulphur for almost its entire drift time towards its iceline (with the exception of the outermost part of the disk in the beginning of the drift, where \ce{H2S} ice dominates). Inside the \ce{SO} iceline ($R\sim15$AU), \ce{SO2} ice dominates, and between 20-40AU \ce{S2} ice peaks (for medium and fast drifts, see Fig. \ref{drifting}) with an abundances of $\sim10^{-6}$ wrt H, and is the second-most abundant sulphur-carrying species. It is noted that observational evidence by, e.g., \citet{kama2019sulphur,legal2021mapsco} suggests that most sulphur in disks may be carried in refractory material. In that context, it is possible that the volatile sulphur species modelled here only account for minor amounts of elemental S.

Figure \ref{pies} shows the ice compositions of the pebbles for the modelling scenarios with drift starting at 200AU. Each row represents a chemical element, and each pie chart shows how much of the total amount of an element is carried in different ice species. The first column shows the compositions at the beginning of drift. The second, third and fourth rows represents the final ice compositions for fast (10kyr), medium (100kyr) and slow drift (1Myr), respectively. These final compositions are taken at a midplane radius of 18.9AU, which is the innermost radius at which none of the considered ices have desorbed yet (except for CO ice, which is also featured in the top left chart).

Reading the pie charts for each element,from left to right, corresponds to considering longer and longer evolution times for the chemistry. Since the pebbles in all scenarios drift at constant radial speeds, the ``Fast'' column features the pebble ice composition after $\sim$9kyr ($R$=18.9AU is $\sim$90\% of the way from 200AU to 1AU on a 10kyr timescale). Similarly, the ``medium'' and ``Slow'' columns features pebbles compositions after $\sim$90kyr and $\sim$0.9Myr of chemical evolution, respectively.

These pie charts show which ice species carry more of one element than other species, and how this changes with drift speed (evolution time). For carbon -and oxygen-carrying species in the top two rows, for example, there is a clear trend from left-to-right that \ce{CO2} ice is increasing in abundance. In the top left chart, \ce{CO2} ice accounts for slightly less that half of total elemental carbon at the beginning of pebble drift. The fraction of elemental carbon in \ce{CO2} then grows, and for slow drift, \ce{CO2} ice accounts for 70\% of all carbon (primarily at the expense of destruction of \ce{CH3OH} ice). For elemental oxygen in the second row, the picture is similar for the \ce{CO2} ice fraction of elemental oxygen (the abundance of \ce{CO2} ice with respect to H$_2$ was counted double in these oxygen pie charts, as \ce{CO2} carries two oxygen atoms per molecule), which grows with drift time. As such, the fraction \ce{H2O}:\ce{CO2}$\sim$5:1 at the start of drift (first column), but changes to \ce{H2O}:\ce{CO2}$\sim$3:2 for slow drift in the last column.

For nitrogen, \ce{NH3} ice remains dominant for different drift timescales, although HCN ice is steadily increasing in abundance with evolution time. \ce{H2S} ice is dominant as a sulphur carrier at the start of drift, and for fast and medium drift. However, for slow drift, SO ice takes over as dominant sulphur carrier, with a tiny fraction of sulphur carried in \ce{SO2} ice. Less abundant volatile species, such as \ce{C2H6}, \ce{C2H4}, \ce{C2H2} and HNC ices are not shown in this work. Out of these, \ce{C2H6} ice reaches an abundance of $4\times10^{-6}$ wrt H at $\sim20$AU for slow drift, but for all other drift timescales, these species remain $<10^{-6}$ wrt H (HNC ice is consistently less abundant than HCN ice, for all drift timescales).

\subsection{Use for modelling planet formation via pebble accretion}

One key goal for the study was to improve the understanding of chemistry in drifting solids, and to provide the planet formation modelling community with insights into these effects. Several efforts over the years have assumed that the chemical abundances assumed initially in models of planet formation would be conserved, even when solids in protoplanetary disks evolve, grow and move. The exception to this is the effects of freeze-out and desorption of molecules, which accounts for the transfers of material between the gas and ice phases.

An important insight from this study is that chemical evolution during drift does not significantly alter the volatile abundances of the dominant carriers of carbon and oxygen. This is the case, especially for slow and medium drift timescales. Quantitatively, the changes to the dominant volatile species, for slow and medium drift is $<20\%$. Following this, when modelling the volatile composition of pebbles that drift from the icy, colder disk and inwards (to be accreted by a forming planet) it is reasonable to assume that the individual abundances of the ices on the pebbles remain largely unchanged during drift, with the exception of desorption of ices into the gas-phase, if icelines are crossed.

\subsection{Does drift origin matter?}

This study has considered drift from two distinct midplane radii. Pebbles originating at each of the two started out with different ice compositions (see Table \ref{initabun}), corresponding to the abundance as its origin radius, as it has been evolved chemically for 100kyr before drifting started. As seen from Fig. \ref{relvals}, the differences in chemical evolution between the two radii is generally between 30\% and a factor of 2 (with the exception of SO and \ce{H2S}). Because the actual drift times for drifting scenarios originating at both radii are the same (the fast drift speed from 128AU is slower than the fast drift speed from 200AU, because the latter needs to drifts almost twice as far in the same time), the changes between the chemical evolution happening from the two radii of origin is likely to be caused by the differences in initial ice abundances for the two cases. These initial differences are, however, not large (see Table \ref{initabun}).

\subsection{Comparison to other work}\label{disccomp}

Recently, \citet{cevallossoto2022} modelled chemical evolution in an evolving disk with drifting pebbles, using the same chemical kinetics code as is used here \citep{walsh2015}. They accounted for pebbles drifting through new material (gas and dust), as they moved through the disk, which is a more realistic approach that the closed parcel-approach that was used here. Their drifting time scales are on the shorter side of what is used here: for their models with stellar accretion rates of $\dot{m}=10^{-8}M_{\odot}$/yr, they find that only pebbles drifting from radii $<10$AU make it to their inner disk region within 100kyr. A radius of 10AU in their Fig. 1 corresponds to a midplane temperature of 60-70K, which is below their \ce{CO2} midplane freeze-out temperature of $\sim80$K. For lower stellar accretion rates of $\dot{m}=10^{-9}M_{\odot}$/yr, their pebble drift speeds are faster (drift times shorter), and so pebbles from the outer disk ($>10$AU) can influence their inner disk on at 100kyr timescale.

\citet{cevallossoto2022} predicts that pebble drift may transport a significant amount of volatiles from the outer disk to the inner disk, with the results that the outer disk loses volatile ices, and the gas abundances around volatile icelines in the inner disk increase. Fig. 2, panels d-f (including pebble drift) in that paper also shows that the abundances of \ce{H2O} ice and \ce{CO2} ice increase with decreasing miplane radius, which is taken to result from chemical evolution in ices during pebble drift. However, a direct comparison between their results for pebble ice compositions during drift, and the results presented in this paper, is not straightforward. This is because their work tracked the abundance of a given ice species at a given midplane radius at a fixed time, where the abundance of an ice species at that radius was made up of the contributions from different pebbles that had drifted from different radii (and had different sizes). In this work, in contrast, the ice composition at a given radius is made up of one (average) pebble of one size drifting from one of two fixed radii, which was tracked \emph{over time}, but assuming different drift time scales.

\subsection{Caveats of model assumptions}
\label{model_caveats}

The modelling setup presented in this paper is a simplification of reality. Most importantly, the pebble drift chemistry modelled here assumes a close parcel of solids and gas, which is being heated up. This means that the drifting pebble is not drifting through new material (material residing further in, in the disk midplane), so there is neither new solids material and new ice compositions being encountered, nor is there new surrounding gas, and possibly higher gas densities around icelines (when a given species desorbs), and increasing gas densities generally moving inwards, as the ices from more and more drifting pebbles desorb and are added to the gas.

It is not straightforward to predict how the effects of a more complete framework including these effects would differ from the effects seen in this paper. However, given that \emph{JWST} is soon to commence operations, it is timely and prudent for the planet formation community, and the astrochemistry community, to come together and develop appropriate modelling frameworks. These need to account for ice-covered pebbles drifting inwards through a disk, where it will encounter gas and solid material that has experienced a different chemical evolution from itself. The frameworks also need to track the effects that pebble drift chemistry has on the ice composition of the pebbles, at it is accreted onto a forming planet.

The opacity in the disk midplane is not evolving along with the dust growth or pebble drift in this work, and the ice mantles on the pebbles are not assumed so thick that parts of them are shielded from UV. Both these aspects should be addressed in future work, in order to make the treatment of the physical and chemical effects more realistic.

The modelling setup presented here only considered one set of initial abundances (per starting radius) for the drifting pebbles. These sets of initial abundances, in turn, were based on global elemental ratios of O:C:N:S of 520:180:60:6.2 \citep[as used in][]{eistrup2016,eistrup2018}. Different choices of global elemental ratios may lead ices of different volatile to evolve differently than seen here. However, following results from \citet{eistrup2018}, one could predict that the same trends for increase and decrease of different volatile would be seen, as the ones seen in this paper, even if starting abundance levels and magnitudes of changes might be different. Future work will explore the impact of different chemical starting conditions.

The drift timescales assumed here do capture the minimum drift timescale as calculated in the Section \ref{methods}, but are generally longer than this. The 100k years and 1M years timescales were chosen to explore chemical evolution in pebbles, assuming that pebbles do not drift inwards at a constant, maximum speed, but also accounting for such effects as dust traps \citep[see, e.g.,][and reference therein]{vandermarel2013,baruteau2019,pinilla2020}. Such traps are generally assumed to either slow down, or halt the drift of solids in a disk. If assuming that a realistic disk does feature more radial structure in the gas and refractory component than what is given by the smooth MMSN power-law surface density profile used in this work, then the drift timescale becomes longer than the minimum timescale calculated earlier. In the context of such longer timescales, this work offers some predictions about the chemical evolution in pebbles, and how this evolution may be different for such a case, as compared to the case for fast drift.

Lastly, this paper has used a single pebble size of $R$=0.6mm for all chemical modelling. This size was motivated by grain growth models by \citet{krijt2016_dustgrowth}, which also informed the accompanying paper \citet{eistrup2022graingrowth}. From pebble drift dynamics there is a general trend that the larger the pebble, the faster it will drift. Adopting one pebble size for three different drift speeds is therefore not necessarily realistic. However, as was shown in \citet{eistrup2022graingrowth}, the larger the pebble (or solid body, generally), the slower the efficiency of ice chemistry on the surfaces of the solid bodies. Therefore, by extension, it can be be derived that if a larger pebble size than 0.6mm had been used for modelling the chemical evolution in the ices over the drift timescales considered here, then this evolution would had been slower than what has been shown in this paper. The results shown here can therefore be considered as extremes for the degree to which chemical evolution can change the icy composition of pebbles. If larger pebbles had been considered for the modelling here, then smaller abundance changes would likely had been observed, but the general trends in the chemical processing (increasing or decreasing abundance for a given volatile ice as a function of pebble drift) would had remained.

\section{Conclusion}

This paper has presented a novel framework for modelling the chemical evolution that ice-covered pebbles of size $R=0.6$mm will experience as they drift inwards from two different radii of origins in a protoplanetary disk, on three different drift timescales. Below are the main conclusions summarized:

\begin{itemize}
    \item Drift speed matters. The slower a pebble drifts, and the longer time it therefore takes it to move from the outer disk to 1AU, the larger are the changes to the abundances of the volatile ice species on the pebble.
    \item Generally, pebble drift chemistry causes \ce{CO2} ice, HCN ice and SO ice to increase in abundance. Elemental oxygen is processed into \ce{CO2} ice (and, to a minor extent, into SO ice), elemental carbon is processed into \ce{CO2} (and, to a minor extent, HCN ice), elemental nitrogen is processed into HCN, and elemental sulphur is processed into SO. For all other volatile ices considered, pebble drift causes a decrease in their abundances.
    \item For medium and fast drift (where the pebble is drifting for 10kyr or 100kyr), the four dominant volatile ices (\ce{H2O}, \ce{CO2}, \ce{CH3OH} and \ce{NH3}) undergo abundance changes of $<20$\%. This means that, assuming such drift time scales, it is reasonable to assume that no chemical evolution happens in the ices, when pebble drift is modelled in relation to planet formation. However, for slower drifts, including possible slowing effects such as pressure bumps, it is possible that chemical evolution during the drift of the pebbles needs to be modelled explicitly.
\end{itemize}

\begin{acknowledgements}
      The authors thank Bertram Bitsch and Dmitry Semenov for interesting discussions that improved the methodology and the quality of the manuscript. Also thank you to the anonymous referee. Research for this paper was supported by the European Research Council under the Horizon 2020 Framework Program via the ERC Advanced Grant Origins 83 24 28.
\end{acknowledgements}
\bibliographystyle{aa} 
\bibliography{refs.bib} 
%
%

\end{document}